\documentclass[12pt]{article}
\usepackage{epsfig}
\usepackage{graphicx}
\usepackage{amsmath}

\begin{document}

 \newcommand{\beq}{\begin{equation}}
\newcommand{\eeq}{\end{equation}}
\newcommand{\bea}{\begin{eqnarray}} 
\newcommand{\eea}{\end{eqnarray}}
\newcommand{\beqn}{\begin{eqnarray}}
\newcommand{\eeqn}{\end{eqnarray}}
\newcommand{\beas}{\begin{eqnarray*}}
\newcommand{\eeas}{\end{eqnarray*}}
\newcommand{\defi}{\stackrel{\rm def}{=}}
\newcommand{\non}{\nonumber}
\newcommand{\bquo}{\begin{quote}}
\newcommand{\enqu}{\end{quote}}
\newcommand{\qt}{\tilde q}
\newcommand{\m}{\tilde m}
\newcommand{\trho}{\tilde{\rho}}
\newcommand{\tn}{\tilde{n}}
\newcommand{\tN}{\tilde N}
\newcommand{\gsim}{\lower.7ex\hbox{$\;\stackrel{\textstyle>}{\sim}\;$}}
\newcommand{\lsim}{\lower.7ex\hbox{$\;\stackrel{\textstyle<}{\sim}\;$}}


\def\de{\partial}
\def\Tr{ \hbox{\rm Tr}}
\def\const{\hbox {\rm const.}}  
\def\o{\over}
\def\im{\hbox{\rm Im}}
\def\re{\hbox{\rm Re}}
\def\bra{\langle}\def\ket{\rangle}
\def\Arg{\hbox {\rm Arg}}
\def\Re{\hbox {\rm Re}}
\def\Im{\hbox {\rm Im}}
\def\diag{\hbox{\rm diag}}


\def\QATOPD#1#2#3#4{{#3 \atopwithdelims#1#2 #4}}
\def\stackunder#1#2{\mathrel{\mathop{#2}\limits_{#1}}}
\def\stackreb#1#2{\mathrel{\mathop{#2}\limits_{#1}}}
\def\Tr{{\rm Tr}}
\def\res{{\rm res}}
\def\Bf#1{\mbox{\boldmath $#1$}}
\def\balpha{{\Bf\alpha}}
\def\bbeta{{\Bf\beta}}
\def\bgamma{{\Bf\gamma}}
\def\bnu{{\Bf\nu}}
\def\bmu{{\Bf\mu}}
\def\bphi{{\Bf\phi}}
\def\bPhi{{\Bf\Phi}}
\def\bomega{{\Bf\omega}}
\def\blambda{{\Bf\lambda}}
\def\brho{{\Bf\rho}}
\def\bsigma{{\bfit\sigma}}
\def\bxi{{\Bf\xi}}
\def\bbeta{{\Bf\eta}}
\def\d{\partial}
\def\der#1#2{\frac{\d{#1}}{\d{#2}}}
\def\Im{{\rm Im}}
\def\Re{{\rm Re}}
\def\rank{{\rm rank}}
\def\diag{{\rm diag}}
\def\2{{1\over 2}}
\def\ntwo{${\mathcal N}=2\;$}
\def\nfour{${\mathcal N}=4\;$}
\def\none{${\mathcal N}=1\;$}
\def\ntwot{${\mathcal N}=(2,2)\;$}
\def\ntwoo{${\mathcal N}=(0,2)\;$}
\def\x{\stackrel{\otimes}{,}}

\def\ba{\beq\new\begin{array}{c}}
\def\ea{\end{array}\eeq}
\def\be{\ba}
\def\ee{\ea}
\def\stackreb#1#2{\mathrel{\mathop{#2}\limits_{#1}}}

\def\Tr{{\rm Tr}}
\newcommand{\cpn}{CP$(N-1)\;$}
\newcommand{\wcpn}{wCP$_{N,\tilde{N}}(N_f-1)\;$}
\newcommand{\wcpd}{wCP$_{\tilde{N},N}(N_f-1)\;$}
\newcommand{\vp}{\varphi}
\newcommand{\pt}{\partial}
\newcommand{\ve}{\varepsilon}
\renewcommand{\theequation}{\thesection.\arabic{equation}}

\setcounter{footnote}0

\vfill

\begin{titlepage}

\begin{flushright}
FTPI-MINN-14/9, UMN-TH-3330/14\\
\end{flushright}

\vspace{1mm}

\begin{center}
{  \Large \bf  
\boldmath{\none  Duality in the Chiral Limit  \\ [2mm] from \ntwo Duality 
}}

\vspace{5mm}

 {\large \bf    M.~Shifman$^{\,a}$ and \bf A.~Yung$^{\,\,a,b}$}
\end {center}

\begin{center}

$^a${\it  William I. Fine Theoretical Physics Institute,
University of Minnesota,
Minneapolis, MN 55455, USA}\\
$^{b}${\it Petersburg Nuclear Physics Institute, Gatchina, St. Petersburg
188300, Russia
}
\end{center}

\vspace{1mm}

\begin{center}
{\large\bf Abstract}
\end{center}

  We study a deformation of \ntwo supersymmetric QCD
with U$(N)$ gauge group and $N_f$ number of quark flavors induced by the mass term $\mu$ 
for the adjoint matter which breaks supersymmetry down to \none QCD. Recently this
deformation was shown to  
lead to a weakly coupled dual theory only in two particular  sets of
vacua: the $r=N$ vacuum and the so-called zero vacua which can be found  at $r< N_f-N$, where  $r$ is 
the number of condensed quarks. For small quark masses and intermediate  values of $\mu$ 
the gauge group of the dual theory is U$(N_f-N)\times U(1)^{2N-N_f}$ where the Abelian sector 
is heavy and can be integrated out. However, at larger values of $\mu$ the Abelian sector enters
the strong coupling regime. We show that the 't Hooft matching conditions in the chiral 
limit require the Seiberg neutral meson field $M$ from this sector to become light. In the $r=N$ vacuum
 $M$ is constructed of a monopole and antimonopole  connected by a confining magnetic strings while
in the zero vacua it is  built of a quark and antiquark  connected by a confining electric strings.

\vspace{2cm}

\end{titlepage}

 \newpage



\section {Introduction }
\label{intro}
\setcounter{equation}{0}

Some time ago we started \cite{Shifman:2007kd} a program of detailing Seiberg's duality \cite{Sdual,IS} in 
${\mathcal N}=1$ theories introducing  masses for the matter fields and exploring diverse discrete vacua
using additional information  (see also \cite{SYdual,SYN1dual} and additional references below) following from the Seiberg-Witten solution  \cite{SW1,SW2} of the ${\mathcal N}=2$ theory. Despite a spectacular  overall progress,
one particular corner of the parameter space, namely its chiral limit,
has not yet been studied, as was noted in   \cite{Chernyak:2013tsa}.
This paper is devoted to thorough studies of the chiral limit, and, thus,  completes the program.
 The picture of the Seiberg duality emerging on the basis of deformations of the ${\mathcal N}=2$ theory is fully self-consistent. It provides a clear-cut understanding of the processes on 
both side of duality.

Seiberg's dual of  \none supersymmetric QCD (SQCD) with the
SU$(N)$ gauge group and $N_f$ quark flavors  is a  theory
with the SU$(\tN)$ gauge group, the same number of dual quarks, plus 
a neutral meson field $M_A^{\,B}$. Here 
\beq
\tN\equiv N_f-N\,.
\label{nnn}
\eeq
Seiberg's duality was generalized to \ntwo supersymmetric QCD deformed by the mass term
$\mu$ for the adjoint matter in the large-$\mu$ limit  \cite{CKM}.
 At large $\mu$ the adjoint matter can be integrated out
leading to a \none QCD-like theory with a quartic superpotential suppressed at large $\mu$ 
\cite{CKM,APS,GVY,GivKut}. This theory has the same number of vacua as that in the original \ntwo QCD 
in the small-$\mu$ limit. These vacua -- the so-called called $r$ vacua -- are characterized by a parameter  $r$, 
the number of condensed (s)quarks in the classical domain of large and generic quark mass parameters
$m_A$ ($A=1,...,N_f$). Clearly, $r$ cannot exceed $N$, the rank of the gauge group.
In the original formulation \cite{Sdual,IS} Seiberg's duality was suggested for the monopole vacua
with $r=0$ (all other vacua become runaway vacua in the limit $\mu\to\infty$). 

Chronologically, the first attempt to obtain Seiberg's duality from $\mu$-deformed \ntwo QCD can be traced back to \cite{APS}.
The dual gauge group SU$(\tN)$ was identified at the root of the baryonic branch.\footnote{It 
corresponds to the $r=N$ quark vacuum in the U$(N)$ version of the theory we consider in this paper.}
However, Seiberg's neutral mesonic fields $M$ were not detected. 

Much later we studied a version of the theory with the U$(N)$ 
gauge group and $(N+1)<N_f<3/2 N$. We demonstrated that
the $\mu$ deformation 
leads to  a weakly coupled dual theory only for two particular  sets of the
vacuum states, namely, in the $r=N$ vacuum and in the so-called zero vacua \cite{SYN1dual,SYzerovac}.
The latter can be found  at $r< \tN$.
 
Both sets of vacua
have  vanishing gaugino condensate in the   limit, in which the values of the quark masses become small.
 In other vacua (the so-called $\Lambda$ vacua) the gaugino condensate is of the order of
 $\mu \Lambda^2_{{\mathcal N}=2}$ where $ \Lambda_{{\mathcal N}=2}$ is the scale of \ntwo QCD. The gaugino condensate
becomes large in the large-$\mu$ limit. Correspondingly, these
 vacua do not have weakly coupled dual description \cite{SYzerovac}.

The gauge group of the dual theory in the $r=N$ and zero vacua is $$ U(\tN)\times U(1)^{N-\tN}\,.$$
For small quark masses and intermediate  values of $\mu$, namely, $$m_A\ll \mu\ll \Lambda_{{\mathcal N}=2}$$ 
the vacuum expectation values (VEVs) of the charged scalar fields in the U$(\tN)$ sector are determined 
by parameters 
\beq
\xi^{\rm small}\sim \mu m
\label{12}
\eeq
 while the VEVs  in the Abelian $U(1)^{N-\tN}$ sector are  determined by
\beq
\xi^{\rm large}\sim \mu \Lambda_{{\mathcal N}=2}\,.
\label{13}
\eeq
Given that $m\ll \Lambda_{{\mathcal N}=2}$ the notation in (\ref{12}) and (\ref{13})
is self-evident.

 The dual theory is infrared free; at intermediate 
values of $\mu$ the both scales $\xi^{\rm small}$ and $\xi^{\rm large}$ are small enough to 
ensure weak coupling. However, the Abelian sector 
is much heavier and thus can be integrated out. Moreover, since  $\sqrt{\xi^{\rm small}}\ll \mu$ in this domain
 the adjoint matter is also heavy and can be integrated out too. This leads to a weakly coupled 
low-energy dual theory
with the U$(\tN)$ Seiberg dual gauge group  and charged light matter \cite{SYN1dual,SYzerovac}. 
Although the correct Seiberg dual gauge group emerges in this setup,  Seiberg's neutral meson $M$ fields 
are still missing. As we will see below, they will show up in the  chiral limit.
 
To this end we make the next step and consider larger values of $\mu$,  
$$\mu\gg \Lambda_{{\mathcal N}=2}\,.$$ In this domain we pass to the chiral limit, or  small quark masses,
keeping the parameter $\xi^{\rm small}$ fixed and small enough to ensure the weak coupling in the U$(\tN)$
sector. At the same time,  the Abelian $U(1)^{N-\tN}$ sector enters a strong coupling regime. Then we use the
't Hooft anomaly matching conditions \cite{Hooft} to show that neutral $M$ mesons coming from this sector
must become light. We find a physical interpretation of the Seiberg $M$ mesons: in the $r=N$ vacuum
 $M$ is constructed of monopole and antimonopole  connected by confining magnetic strings, while
in the zero vacua 
 $M$  is   constructed of quark and antiquark  connected by confining electric strings.
The match of our dual description in these sets of vacua with Seiberg's dual theory becomes complete.

In the first part of the paper (Secs. \ref{rNsmallmu} and \ref{intermediatemu}) we briefly summarize our 
previous results on $r$ duality outside the chiral limit, emphasizing
 its peculiarities, such as ``instead-of-confinement" mechanism. In Sec. \ref{largemu} we pass to the exploration of the chiral limit, and discover that the neutral Seiberg $M^{\,B}_A$ mesons
show up in the light sector. Thus, $r$ duality proves to be completely woven in the fabric of Seiberg's duality.

The paper is organized as follows. In Sec.~\ref{rNsmallmu} we review duality  and ``instead-of-confinement''
mechanism in $r=N$ vacuum in \ntwo limit of small $\mu$. In Sec.~\ref{intermediatemu}
 we review the dual theory at 
intermediate $\mu$.
Next in Sec.~\ref{largemu} we consider large $\mu$ and use anomaly matching conditions to show that
monopole-antimonopole stringy mesons originating from the Abelian  $U(1)^{N-\tN}$ sector of the theory
should become light. We also present the dual  low energy theory in this region and discuss it 
mass spectrum.
In Sec.~\ref{zerovacua} we review the low energy description in $r$-vacua with $r<N_f/2$ at small $\mu$. 
In  Sec.~\ref{0vacuamu} we consider subset of these vacua, namely zero  vacua at 
intermediate  and large $\mu$ 
 and show that stringy quark-antiquark mesonic states 
should become light as we increase $\mu$. Sec.~\ref{conclusion} contains our summary and conclusions.

\section {Duality in the \boldmath{$r=N$} vacuum at small \boldmath{$\mu$}}
\label{rNsmallmu}
\setcounter{equation}{0}

In this section we briefly review non-Abelian duality in the $r=N$ vacua at small $\mu$ established
 in \cite{SYdual,SYtorkink}.
The gauge symmetry of our basic model is 
$$U(N)=SU(N)\times U(1)\,.$$ In the absence
of  deformation the model under consideration is \ntwo  SQCD
 with $N_f$ massive quark hypermultiplets. 
 We assume that
$N_f>N+1$ but $N_f<\frac32 N$. 
The latter inequality ensures  that the dual theory can be infrared free.

Our basic  theory is described in detail in our previous papers (e.g. \cite{SYmon,SYfstr}; see also
 the reviews in \cite{SYrev}).
The field content is as follows. The \ntwo vector multiplet
consists of the  U(1)
gauge field $A_{\mu}$ and the SU$(N)$  gauge field $A^a_{\mu}$,
where $a=1,..., N^2-1$, and their Weyl fermion superpartners plus
complex scalar fields $a$, and $a^a$ and their Weyl superpartners, respectively.

As for the matter sector,
the $N_f$ quark multiplets of  the U$(N)$ theory consist
of   the complex scalar fields
$q^{kA}$ and $\tilde{q}_{Ak}$ (squarks) and
their   fermion superpartners --- all in the fundamental representation of 
the SU$(N)$ gauge group.
Here $k=1,..., N$ is the color index
while $A$ is the flavor index, $A=1,..., N_f$. We will treat $q^{kA}$ and $\tilde{q}_{Ak}$
as rectangular matrices with $N$ rows and $N_f$ columns. 

In addition, we introduce the mass term $\mu$ 
for the adjoint matter breaking \ntwo supersymmetry down to \none. 
This deformation term
\beq
{\mathcal W}_{{\rm def}}=
  \mu\,{\rm Tr}\,\Phi^2, \qquad \Phi\equiv\frac12\, {\mathcal A} + T^a\, {\mathcal A}^a
\label{msuperpotbr}
\eeq
does not break \ntwo supersymmetry in the small-$\mu$ limit,  see \cite{HSZ,VY,SYmon}.
At large $\mu$ this theory obviously flows to \none\!. The fields  ${\mathcal A}$ and ${\mathcal A}^a$ 
in Eq.~(\ref{msuperpotbr}) are  chiral superfields, the ${\mathcal N}=2$
superpartners of the U(1) and SU($N$) gauge bosons.

\subsection{The \boldmath{$r=N$} vacuum at large \boldmath{$\xi$}}

This theory has a set of $r$ vacua, where  $r$ is 
the number of condensed (s)quarks in the classical domain of large generic quark masses 
$m_A$ ($A=1,...,N_f$, and $r\le N$). In the first part of this paper we consider the $r=N$ vacua
(for a review see \cite{SYrev}). These
 vacua have  the maximal possible number of condensed quarks, $r=N$. Moreover,
  the gauge group U$(N)$ is completely 
Higgsed in these vacua, and, as a result, 
they support non-Abelian strings \cite{HT1,ABEKY,SYmon,HT2}. The occurrence of 
these strings ensures 
confinement of monopoles in these vacua. 

First, we  will assume that $\mu$ is small, much smaller than
the quark masses
\beq
|\mu| \ll | m_A |, \qquad A=1, ..., N_f\,.
\label{smallmu}
\eeq

 In the quasiclassical region of large quark masses scalar quarks develop VEVs triggered by the 
deformation parameter $\mu$. They are given by
\beqn
\langle q^{kA}\rangle &=& \langle\bar{\tilde{q}}^{kA}\rangle=\frac1{\sqrt{2}}\,
\left(
\begin{array}{cccccc}
\sqrt{\xi_1} & \ldots & 0 & 0 & \ldots & 0\\
\ldots & \ldots & \ldots  & \ldots & \ldots & \ldots\\
0 & \ldots & \sqrt{\xi_N} & 0 & \ldots & 0\\
\end{array}
\right),
\nonumber\\[4mm]
k&=&1,..., N\,,\qquad A=1,...,N_f\, ,
\label{qvev}
\eeqn
where we present the quark fields as  matrices in the color ($k$) and flavor ($A$) indices,
 while parameters $\xi$ are given in the quasiclassical
approximation by 
\beq
\xi_P \approx 2\;\mu m_P,
\qquad P=1,...,N.
\label{xis}
\eeq

The quark condensate (\ref{qvev}) result in  the spontaneous
breaking of both gauge and flavor symmetries.
A diagonal global SU$(N)$ combining the gauge SU$(N)$ and an
SU$(N)$ subgroup of the flavor SU$(N_f)$
group survives in the limit of (almost) equal quark masses. 
This is color-flavor locking. 

Thus, the unbroken global symmetry 
is as follows: 
\beq
  {\rm SU}(N)_{C+F}\times  {\rm SU}(\tN)\times {\rm U}(1)\,.
\label{c+f}
\eeq
Here SU$(N)_{C+F}$ is a global unbroken color-flavor rotation, which involves the
first $N$ flavors, while the SU$(\tN)$ factor stands for the flavor rotation of the 
$\tN$ quarks.

The presence of the global SU$(N)_{C+F}$ group is the reason for
formation of the non-Abelian strings \cite{HT1,ABEKY,SYmon,HT2,SYfstr}.
At small $\mu$ these strings are BPS-saturated \cite{HSZ,VY} and their
tensions  are determined  by 
the parameters $\xi_P$ \cite{SYfstr}, see (\ref{xis}),
\beq
T_P=2\pi|\xi_P|\, , \qquad P=1,...,N.
\label{ten}
\eeq
These string confine monopoles. In fact, in the U$(N)$ theories confined  elementary monopoles 
are junctions of two ``neighboring'' $P$-th and $(P+1)$-th strings, see \cite{SYrev} for a review.

Now, let us briefly discuss the perturbative excitation spectrum. 
Since
both U(1) and SU($N$) gauge groups are broken by the squark condensation, all
gauge bosons become massive.

To the leading order in $\mu$, \ntwo supersymmetry is not broken.  In fact, with 
nonvanishing $\xi_P$'s (see Eq.~(\ref{xis})), both the quarks and adjoint scalars  
combine  with the gauge bosons to form long \ntwo supermultiplets \cite{VY}. 
In the equal  mass limit  $\xi_P\equiv\xi\,,$  and all states come in 
representations of the unbroken global
 group (\ref{c+f}), namely, in the singlet and adjoint representations
of SU$(N)_{C+F}$,
\beq
(1,\, 1), \quad (N^2-1,\, 1),
\label{onep}
\eeq
 and in the bifundamental representations
\beq
 \quad (\bar{N},\, \tN), \quad
(N,\, \bar{\tN})\,.
\label{twop}
\eeq
The representations in (\ref{onep}) and (\ref{twop})  are marked with respect to two 
non-Abelian factors in (\ref{c+f}). The singlet and adjoint fields are (i) the gauge bosons, and
(ii) the first $N$ flavors of squarks $q^{kP}$ ($P=1,...,N$), together with their 
fermion superpartners.
The bifundamental fields are the quarks $q^{kK}$ with $K=N+1,...,N_f$.
Quarks transform in the two-index representations of the global
group (\ref{c+f}) due to the color-flavor locking. 

The above quasiclassical analysis is valid if the theory is at weak coupling. 
From (\ref{qvev}) we see that the weak coupling condition is 
\beq
\sqrt{\xi} \sim \sqrt{\mu m}\gg\Lambda_{{\mathcal N}=2}\,,
\label{weakcoup}
\eeq
where  we assume all quark masses 
to be of the same order $m_A\sim m$. This condition means that 
the quark masses are large enough to compensate
the smallness of $\mu$.

\subsection{\boldmath{$r$} Dual theory}
\label{vacjumpN}

Now we will relax the condition (\ref{weakcoup}) and pass
to the strong coupling domain at 
\beq
|\sqrt{\xi_P}|\ll \Lambda_{{\mathcal N}=2}\,, \qquad | m_{A}|\ll \Lambda_{{\mathcal N}=2}\,,
\label{strcoup}
\eeq
still keeping $\mu$ small.
 
As was shown in \cite{SYdual,SYN1dual}  in the $r=N$ vacuum \ntwo QCD undergoes a
crossover transition as the value of $\xi$ decreases.
The  domain (\ref{strcoup}) 
can be described in terms of weakly coupled (infrared free) $r$-dual theory 
with  the  gauge group
\beq
{\rm U}(\tN)\times {\rm U}(1)^{N-\tN}\,,
\label{dualgaugegroup}
\eeq
 and $N_f$ flavors of light quark-like dyons.\footnote{ Previously the SU$(\tN)$
 gauge group  was identified  \cite{APS}   
at the root of the baryonic Higgs branch in the \ntwo supersymmetric  SU($N$) Yang--Mills 
theory with massless quarks and vanishing  $\xi$ parameters.}
Note, that we call our 
dual theory the ``$r$ dual'' because \ntwo duality described here can be generalized to other
$r$ vacua with $r>N_f/2$.  This leads to a theory with the dual gauge group U$(N_f-r)\times$U(1)$^{N-N_f+r}$
\cite{SYrvacua}. However, deformation of these $r$ dual theories to 
\none theory at larger $\mu$ can be performed within the weak coupling regime only in the $r=N$ 
vacuum \cite{SYzerovac}, which we discuss here.

\vspace{2mm}

The light dyons $D^{lA}$ 
($l=1, ... ,\tN$ and $A=1, ... , N_f$) are in 
the fundamental representation of the gauge group
SU$(\tN)$ and are charged under the Abelian factors indicated in Eq.~(\ref{dualgaugegroup}).
 In addition, there are  $(N-\tN)$ 
light dyons $D^J$ ($J=\tN+1, ... , N$), neutral under 
the SU$(\tN)$ group, but charged under the
U(1) factors. 

The color charges of all these dyons are identical to those of quarks.\footnote{Because of monodromies \cite{SW1,SW2,BF} the quarks pick up root-like color-magnetic 
charges in addition to their weight-like color-electric charges at strong coupling \cite{SYdual}.}
This is the reason why we call them quark-like dyons. However, 
these dyons are not quarks \cite{SYdual}. As we will review below they belong to a
different representation of the global color-flavor locked group.
Most importantly, condensation of these dyons still 
leads to confinement of monopoles.

The dyon condensates have the form \cite{SYfstr,SYN1dual}:
\beqn
\!\!\!\!
\langle D^{lA}\rangle \!\!\! \!& =& \langle \bar{\tilde{D}}^{lA}\rangle \! =
\frac1{\sqrt{2}}\,\left(
\begin{array}{cccccc}
0 & \ldots & 0 & \sqrt{\xi_{1}} & \ldots & 0\\
\ldots & \ldots & \ldots  & \ldots & \ldots & \ldots\\
0 & \ldots & 0 & 0 & \ldots & \sqrt{\xi_{\tN}}\\
\end{array}
\right)\!,
\label{Dvev}
\\[4mm]
\langle D^{J}\rangle &=& \langle\bar{\tilde{D}}^{J}\rangle=\sqrt{\frac{\xi_J}{2}}, 
\qquad J=(\tN +1), ... , N\,.
\label{adiiz}
\eeqn
The  important feature apparent in (\ref{Dvev}), as compared to the squark VEVs  in the 
original theory (\ref{qvev}),  is a ``vacuum leap'' \cite{SYdual}.
Namely, if we pick up the vacuum with nonvanishing VEVs of the  first $N$ quark flavors
in the original theory at large $\xi$,  and then reduce $\xi$ below 
$\Lambda_{{\mathcal N}=2}$, 
the system goes through a crossover transition and ends up in the vacuum 
of the  $r$-dual theory with the dual gauge group (\ref{dualgaugegroup}) and 
nonvanishing VEVs of $\tN $ last dyons (plus VEVs of $(N-\tN)$ dyons 
that are the SU$ (\tN)$ singlets).

The parameters $\xi_P$  in (\ref{Dvev}) and (\ref{adiiz}) are determined by the 
quantum version of the classical expressions
(\ref{xis}) \cite{SYfstr}.   
They can be expressed   in  terms of the roots of the Seiberg--Witten curve \cite{SW1,SW2}. The
 Seiberg--Witten curve in our theory has the form \cite{APS}
\beq
y^2= \prod_{P=1}^{N} (x-\phi_P)^2 -
4\left(\frac{\Lambda_{{\mathcal N}=2}}{\sqrt{2}}\right)^{N-\tN}\, \,\,\prod_{A=1}^{N_f} 
\left(x+\frac{m_A}{\sqrt{2}}\right),
\label{curve}
\eeq
where $\phi_P$ are gauge invariant parameters on the Coulomb branch.

In the $r=N$ vacuum the curve (\ref{curve}) has $N$ double roots and reduces to
\beq
y^2= \prod_{P=1}^{N} (x-e_P)^2.
\label{rNcurve}
\eeq
This reflects the condensation of $N$ quarks.
Quasiclassically, at large masses, $e_P$'s are given  by the
mass parameters, $\sqrt{2}e_P\approx -m_P$ ($P=1, ... , N$).

The dyon condensates (\ref{Dvev}) at small masses in the $r=N$ vacuum are determined by 
\cite{SYfstr,SYN1dual}
\beq
\xi_P=-2\sqrt{2}\,\mu\,e_P\,.
\label{xirN}
\eeq

As long as we keep $\xi_P$ and masses small enough (i.e. in the domain (\ref{strcoup}))
the coupling constants of the
infrared-free $r$-dual theory (frozen at the scale of the dyon VEVs) are small:
the $r$-dual theory is at weak coupling.

At small masses, in the region  (\ref{strcoup}), the double roots of the Seiberg--Witten
 curve are  
\beqn
&&
\sqrt{2}e_I = -m_{I+N}, \qquad 
\sqrt{2}e_J = \Lambda_{{\mathcal N}=2}\,\exp{\left(\frac{2\pi i}{N-\tN}J\right)},
\nonumber\\
&&
I=1, ... ,\tN\,\,\,\, {\rm  and} \,\,\,\, J=(\tN+1), ... , N\,.
\label{roots}
\eeqn
In particular, the $\tN$ first roots are determined by the masses of the 
 last $\tN$ quarks --- a reflection of the fact that the 
non-Abelian sector of the dual theory is infrared free and is at weak coupling
in the domain (\ref{strcoup}). 

\subsection{``Instead-of-confinement'' mechanism}
\label{instoc}

Now, we will consider the limit of almost equal quark masses.
Both, the gauge group and the global flavor SU($N_f$) group, are
broken in the vacuum. However, the form of the dyon VEVs in (\ref{Dvev}) shows that the $r$-dual theory 
is also in the color-flavor locked phase. 
 Namely, the  unbroken  global group of the dual
theory is 
\beq
 {\rm SU}(N)\times  {\rm SU}(\tN)_{C+F}\times {\rm U}(1)\,,
\label{c+fd}
\eeq
where this time the SU$(\tN)$ global group arises from   color-flavor locking.

 In much the same way as 
in the original theory, the presence of the global SU$(\tN)_{C+F}$ symmetry
is the  reason behind formation of the non-Abelian strings. Their tensions 
are still given by Eq.~(\ref{ten}),
where the parameters $\xi_P$ are determined by (\ref{xirN}) \cite{SYfstr,SYN1dual}.
These strings still confine monopoles \cite{SYdual}.\footnote{An explanatory remark regarding
our terminology is in order. Strictly speaking, the  dyons carrying root-like electric charges
are confined as well. We refer to all such states collectively as to
``monopoles.'' This is to avoid confusion with the quark-like dyons which appear in Eqs.~(\ref{Dvev}) and
(\ref{adiiz}). The
latter dyons carry weight-like electric charges. As was already mentioned, their color charges 
are identical to those of quarks, see \cite{SYdual} for further details.}
  
In the equal-mass limit
the global unbroken symmetry (\ref{c+fd}) of the dual theory at small
$\xi$ coincides with the global group (\ref{c+f}) of the original theory in the
$r=N$ vacuum at large $\xi$.  
However,  this global symmetry is realized in two very distinct ways in the dual pair at hand.
As was already mentioned, the quarks and U($N$) gauge bosons of the original theory at large $\xi$
come in the following representations 
 of the global group (\ref{c+f}):
 $$
 (1,1), \,\, (N^2-1,1), \,\, (\bar{N},\tN), \,\,{\rm and} \,\, (N,\bar{\tN})\,.
 $$
At the same time,  the dyons and U($\tN$) gauge 
bosons of the $r$-dual theory form 
\beq
(1,1),\,\, (1,\tN^2-1),\,\, (N,\bar{\tN}), \,\, {\rm and}\,\,
(\bar{N},\tN)
\label{represd}
\eeq 
representations of (\ref{c+fd}). We see that the
adjoint representations of the $(C+F)$
subgroup are different in two theories. 
     
The quarks and gauge bosons
which form the  adjoint $(N^2-1)$ representation  
of SU($N$) at large $\xi$ and the quark-like dyons and dual gauge bosons which 
form the  adjoint $(\tN^2-1)$ 
representation  of SU($\tN$) at small $\xi$ are, in fact, {\em distinct} states \cite{SYdual}.

Thus, the quark-like dyons are not quarks. 
At large $\xi$ they are heavy solitonic states. However below the crossover
at small $\xi$ they become light and form the fundamental ``elementary" states $D^{lA}$ of the $r$-dual theory.
And {\em vice versa}, quarks are light at large $\xi$ but become heavy below the crossover.

This raises the question: what exactly happens with quarks when we reduce $\xi$? 

They are in the ``instead-of-confinement'' phase. The
Higgs-screened quarks and gauge bosons 
at small $\xi$  decay into the monopole-antimonopole 
pairs on the curves of marginal stability (the so-called wall crossing) \cite{SYdual,SYtorkink}. 
The general rule is that the only states that exist at strong coupling inside the curves of marginal stability 
are those which can become massless on the Coulomb branch
\cite{SW1,SW2,BF}. For the $r$-dual theory these are light dyons shown in Eq.~(\ref{Dvev}),
gauge bosons of the dual gauge group and monopoles.
 
\begin{figure}
\epsfxsize=6cm
\centerline{\epsfbox{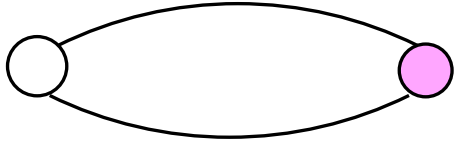}}
\caption{\small Meson formed by a monopole-antimonopole pair connected by two strings.
Open and closed circles denote the monopole and antimonopole, respectively.}
\label{figmeson}
\end{figure}

 At small nonvanishing values of $\xi$ the
monopoles and antimonopoles produced in the decay process of the adjoint $(N^2-1,1)$ states
cannot escape from
each other and fly  to opposite infinities 
because they are confined. Therefore, the (screened) quarks and  gauge bosons 
evolve into stringy mesons  (in the strong coupling domain of small  $\xi$) shown in  Fig.~\ref{figmeson},
namely monopole-antimonopole
pairs connected  by two strings \cite{SYdual,SYN1dual}. 

The flavor quantum numbers of stringy  monopole-antimonopole mesons were studied in 
\cite{SYtorkink} in the framework of an appropriate two dimensional   $CP(N-1)$  model which describes 
world sheet dynamics of the non-Abelian strings \cite{HT1,ABEKY,SYmon,HT2}. In particular, confined monopoles 
are seen as kinks in this world sheet theory. If two strings in Fig.~\ref{figmeson} are ``neighboring''
strings $P$ and $P+1$ ($P=1,...,(N-1)$), each meson is in the two-index representation $M_A^B(P,P+1)$
of the flavor group, where the flavor indices are 
$A,B =1,..., N_f$. It splits into singlet, adjoint and bifundamental representations of 
the global unbroken group (\ref{c+fd}). In particular, at small $\xi$ the adjoint representation
of SU$(N)$  contains former (screened) quarks and gauge bosons of the original theory. 

Masses of these stringy mesons are determined by string tensions  given by the parameters $\xi_P$,
$\xi_{P+1}$, see (\ref{xirN}) and (\ref{roots}). In particular, in the $r$-dual theory the tensions
of  $\tN$ non-Abelian strings from the U$(\tN)$ sector are light, of the order of $\xi^{\rm small}\sim \mu m$,
 while the 
tensions of $(N-\tN)$ ``Abelian'' strings from U$(1)^{N-\tN}$ sector are much heavier, of the order of 
 $\xi^{\rm large}\sim \mu  \Lambda_{{\mathcal N}=2}$. The majority  of stringy mesons  are unstable
 and decay into each other or into the ``elementary" states
(\ref{represd}) of the $r$-dual theory, the dyons and  gauge  bosons. For example, the  mesons $M_A^B(P,P+1)$
which form 
 representations (\ref{represd}) can decay into elementary states with the same quantum numbers
 \cite{SYdual,SYtorkink}.

\section {Intermediate \boldmath{$\mu$}}
\label{intermediatemu}
\setcounter{equation}{0}

In this section we will discuss what happens to the $r$-dual theory in the $r=N$ vacuum 
described above once we increase $\mu$ to 
intermediate values, which are large enough to decouple the adjoint matter \cite{SYN1dual,SYrvacua}.
We also discuss the relation of our dual theory to the Seiberg's dual.

\subsection{Emergence of the \boldmath{$U(\tN)$} gauge group}
\label{UtN}

Combining Eqs.~(\ref{Dvev}), (\ref{adiiz}), (\ref{xirN}) and (\ref{roots}) 
we see that the VEVs of the non-Abelian
dyons $D^{lA}$ are determined by 
\beq
\sqrt{\xi^{\rm small}}\sim\sqrt{\mu m} 
\label{xismall}
\eeq
and are much smaller 
than the VEVs of the Abelian dyons $D^{J}$ in the domain
(\ref{strcoup}).  The latter are of the order of 
\beq
\sqrt{\xi^{\rm large}}\sim \sqrt{\mu \Lambda_{{\mathcal N}=2}}. 
\label{xilarge}
\eeq
This circumstance is most crucial. It  allows us to  increase $\mu$
and decouple the adjoint fields without violating  the weak coupling condition in the 
dual theory \cite{SYN1dual}.

Let us uplift $\mu$ to the intermediate domain
\beq
|\mu| \gg |m_A |, \qquad A=1, ... , N_f\, , \qquad \mu\ll \Lambda_{{\mathcal N}=2} .
\label{muintermed}
\eeq
The VEVs of the Abelian dyons (\ref{adiiz}) are large. This makes U(1) gauge 
fields of the dual group 
(\ref{dualgaugegroup}) heavy. Decoupling these gauge factors, together with 
the adjoint matter and 
the Abelian dyons themselves, we obtain the low-energy theory with the
\beq
U(\tN)
\label{dualgglmurN}
\eeq
gauge fields and the following  set of non-Abelian dyons: $D^{lA}$ ($l=1, ... , \tN$, $\,A=1, ... , N_f$).
The superpotential 
for $D^{lA}$ has the form \cite{SYN1dual}
\beq
{\mathcal W} = -\frac1{2\mu}\,
(\tilde{D}_A D^B)(\tilde{D}_B D^A)  
+m_A\,(\tilde{D}_A D^A)\,,
\label{superpotd}
\eeq
where the color indices are contracted inside each parentheses.
Minimization of this superpotential leads to the VEVs (\ref{Dvev}) of non-Abelian dyons
determined by  $\xi^{\rm small}$, see (\ref{roots}).

Below the scale $\mu$ our theory becomes dual to \none SQCD with the scale
\beq
\tilde{\Lambda}_{{\mathcal N}=1}^{N-2\tN}= \frac{\Lambda_{{\mathcal N}=2}^{N-\tN}}{\mu^{\tN}}\,.
\label{tildeL}
\eeq
In order to keep this infrared-free theory in the weak coupling 
regime we impose that 
\beq
|\sqrt{\mu m}| \ll \tilde{\Lambda}_{{\mathcal N}=1}\,.
\label{wcdual}
\eeq
This means that at large $\mu$ we must keep the quark masses sufficiently small. 

Let us briefly summarize the mass spectrum of our U$(\tN)$ $r$-dual theory at intermediate $\mu$ 
\cite{SYN1dual}. The lightest states are $4N\tN$ bifundamental dyons 
(we count real bosonic degrees of freedom). Their masses are of the order of quark mass differences $(m_A-m_B)$.
A half of dyons, namely, $2\tN^2$, from singlet and adjoint representations of SU$(\tN)$ are also light
with masses of the order of $m\sim m_A$. Another $\tN^2$ dyonic states become scalar superpartners for the massive
gauge bosons of the $U(\tN)$ gauge group (altogether $4\tN^2$ states). These are much heavier, with masses
of the order of $\tilde{g}\sqrt{\xi^{\rm small}}$, where $\tilde{g}$ is the gauge coupling 
constant of the $r$-dual theory. On top of that we have stringy monopole-antimonopole
mesons (see Fig.~\ref{figmeson}) $M_A^B(P,P+1)$, where  $P=1,...,(\tN-1)$, while $A,B=1,...,N_f$.
Their masses are  of the order of $\sqrt{\xi^{\rm small}}$; they are
determined by tensions of light non-Abelian strings. 

Note that in the intermediate domain of  $\mu$  (\ref{muintermed})  we assume that 
$\mu\ll \Lambda_{{\mathcal N}=2}$. This condition ensures that the heavy Abelian $U(1)^{(N-\tN)}$
sector is at weak coupling too and really heavy. At weak coupling the masses of the states
in this sector can be  determined in the quasiclassical approximation. They are of the order of 
 $g_{U(1)}\sqrt{\xi^{\rm large}}$ for ``elementary"  states,
where $g_{U(1)}$ are couplings in the $U(1)$ factors,  and are of the order of $\sqrt{\xi^{\rm large}}$ for stringy mesons
$M_A^B(P,P+1)$ with $P=\tN,...,(N-1)$. 

If we relax the  condition $\mu\ll \Lambda_{{\mathcal N}=2}$ 
this sector enters a strong coupling regime
and certain states could in principle become light and couple to our low-energy $U(\tN)$ theory.
We will see in the next section that this is exactly what happens at larger values of $\mu$ and
is, in fact, required by the 't Hooft anomaly matching \cite{Hooft}.

\subsection{Connection to Seiberg's duality}
\label{Sconnection}

The gauge group of our $r$-dual theory is U$(\tN)$, the same as the gauge group of the Seiberg's dual
theory \cite{Sdual,IS}.
 This suggests that there should be a close relation between two duals. For intermediate values of $\mu$
this relation was found in \cite{SYvsS,SYzerovac}.

Originally Seiberg's duality was formulated for \none SQCD which in our set-up  corresponds to the limit $\mu\to \infty$.
Therefore, in the original formulation Seiberg's duality referred to the monopole vacua with $r=0$.
Other vacua, with $r\ne 0$, have condensates of $r$ quark flavors  $\langle \tilde{q}q\rangle_{A} \sim \mu m_A$
and, therefore, disappear in the limit  $\mu\to \infty$: they become runaway vacua. 
However, as was already mentioned in Sec. \ref{intro},  Seiberg's duality can be generalized to the 
$\mu$-deformed \ntwo QCD  \cite{CKM,GivKut}. At large $\mu$, $\mu$-deformed \ntwo QCD flows to \none QCD with an
additional quartic quark superpotential. This theory has all $r$ vacua which were present in original \ntwo
QCD in the small-$\mu$ limit. 
The generalized Seiberg's dual theory for the  $\mu$-deformed U$(N)$
\ntwo SQCD at large but finite $\mu$  has the gauge group U$(\tN)$,
 $N_f$ flavors of Seiberg's ``dual quarks'' $h^{lA}$ ($l=1,...,\tN$ and $A=1,...,N_f$) and the  superpotential
\beq
{\cal W}_{S}= -\frac{\kappa^2}{2\mu}\,{\rm Tr}\,(M^2) + \kappa \,m_A\, M_A^A +\tilde{h}_{Al}h^{lB}\,M_B^A,
\label{Ssup}
\eeq
where $M_A^B$ is the Seiberg's neutral mesonic field defined as
\beq
(\tilde{q}_A q^B)=\kappa\,M_A^B.
\label{M}
\eeq
Here $\kappa$ is a parameter of dimension of mass needed to formulate Seiberg's duality \cite{Sdual,IS}.
Two last terms in (\ref{Ssup}) were originally suggested by Seiberg, while the first term is a generalization
to finite $\mu$ which originates from the quartic quark potential \cite{CKM,GivKut}.

Now let us assume the fields $M_A^B$ to be heavy and integrate them out. This implies that $\kappa$ is 
large.
Integrating out the $M$ fields in (\ref{Ssup}) we get
\beq
{\cal W}_{S}^{\rm LE} = \frac{\mu}{2\kappa^2}\, (\tilde{h}_{A}h^{B})(\tilde{h}_{B}h^{A}) + 
\frac{\mu}{\kappa}\,m_A\,(\tilde{h}_{A}h^{A})\,.
\label{SLEsup}
\eeq
The  change of variables
\beq
D^{lA}=\sqrt{-\frac{\mu}{\kappa}}\, h^{lA}, \qquad l=1,...,\tN, \qquad A=1,...,N_f
\label{change}
\eeq
brings this superpotential to the form 
\beq
{\cal W}_{S}^{\rm LE} = \frac1{2\mu}\,
(\tilde{D}_A D^B)(\tilde{D}_B D^A)  
-m_A\,(\tilde{D}_A D^A)\,.
\label{S=SY}
\eeq
We see that (up to a sign) this superpotential coincides   with the superpotential of our $r$-dual theory
(\ref{superpotd}). As was already mentioned, the
 dual gauge groups  also coincide for Seiberg's and $r$-dual theories in the $r=N$ vacuum. Note, that 
the kinetic 
terms are not known in the Seiberg's dual theory; thus, normalization of the $h$ fields is  not   fixed.

We see that the $r$-dual and Seiberg's dual theories match. However, it seems that this match is not 
complete. The mesonic field $M_A^B$ is supposed to be light in the Seiberg duality.  

It seems, there is
no apparent candidate for a light neutral field with these flavor quantum numbers in the $r$-dual theory.
 Moreover,
the match outlined above assumes that the $M$ field is heavy and can be integrated out. 

In principle, there are 
candidates for the Seiberg $M$ field with correct flavor quantum numbers in the $r$-dual theory. These are 
the monopole-antimonopole stringy mesons  $M_A^B(P,P+1)$ from the  Abelian sector with  $P=\tN,...,(N-1)$.
They could produce the Seiberg  $M$ field.  But ...

At intermediate $\mu$ (\ref{muintermed}) the
U$(1)^{(N-\tN)}$  Abelian sector  is at weak coupling. This ensures that the
masses of the Abelian $M_A^B(P,P+1)$ mesons can be determined quasiclassically.
 As was discussed in  
Sec. \ref{UtN},  they are of the order of  $\sqrt{\xi^{\rm large}}$  and cannot possibly become light.
We will come back to this issue in Sec.~\ref{interpM}.

The resolution of this puzzle is that Seiberg's duality refers to a much larger values of $\mu$
than those given by the upper bound in (\ref{muintermed}). In fact, the generalized Seiberg duality  
assumes that 
\beq
\mu\gg \Lambda_{{\mathcal N}=1}
\label{originallylargemu}
\eeq
where  $\Lambda_{{\mathcal N}=1}$ is the scale of the original \none QCD
\beq
\Lambda_{{\mathcal N}=1}^{2N-\tN}= \mu^{N}\,\Lambda_{{\mathcal N}=2}^{N-\tN}\,.
\label{Lambda}
\eeq
The domain (\ref{originallylargemu}) is  above the intermediate-$\mu$ domain considered in 
this section.

This leads us to the conclusion that at intermediate $\mu$ we have a perfect match between the $r$-dual and 
Seiberg's
dual theories. In this domain the Seiberg $M$ meson is heavy and should be integrated out implying the 
superpotential (\ref{S=SY}) which agrees with the superpotential (\ref{superpotd}) obtained in the $r$-dual 
theory.

This match, together with the identification (\ref{change}), 
reveals the physical nature of Seiberg's ``dual quarks.''
They are not monopoles as naive duality suggests. Instead,
 they are quark-like dyons appearing in the $r$-dual theory below the crossover. Their condensation leads to 
confinement of monopoles and the ``instead-of-confinement'' phase \cite{SYrvacua} for quarks and gauge bosons of the 
original theory.

\section {Large \boldmath{$\mu$}}
\label{largemu}
\setcounter{equation}{0}

Now we turn to the large-$\mu$ domain. Increasing $\mu$ we
simultaneously reduce $m$ keeping $\xi^{\rm small}$  sufficiently small, see (\ref{wcdual}).
Namely, we assume
\beq
\xi^{\rm small}\sim \mu m \ll  \tilde{\Lambda}_{{\mathcal N}=1},
\qquad  \mu \gg \Lambda_{{\mathcal N}=1}.
\label{mularge}
\eeq
This ensures that our low-energy U$(\tN)$ $r$-dual theory is at weak coupling. However, the Abelian
U$(1)^{(N-\tN)}$ sector ultimately enters the strong coupling regime. As was already mentioned, we loose
analytic control over this sector and, in particular, certain states can  become light and couple to 
our low-energy U$(\tN)$ theory.
Below we will show that this indeed happens, as required by the 't Hooft anomaly matching. 

The anomaly matching was previously analyzed in \cite{Sdual} as a basis for the very formulation of 
the Seiberg duality. In particular, the anomaly matching requires to have light neutral meson $M$ field
in the dual theory. Without Seiberg's $M$ meson the anomalies do not match. A  novelty of our
discussion  in this section is that we have a symmetry breaking in the $r$-dual theory at 
the scale $\sqrt{\xi^{\rm small}}$ and have to match anomalies at energies above and below this scale.
This leads to a rather restrictive bound for the $M$-meson mass. Also, since we $\mu$-deform our
$r$-dual theory and start from a well understood \ntwo limit, we can reveal a physical interpretation for the $M$ meson.

\subsection{Anomaly matching}

The limit (\ref{mularge}) ensures that the quark masses are rather small. They are the smallest parameters of the
theory. Thus, we are in the chiral limit. Above the scale $m$ the global group of our theory before the
symmetry breaking includes independent left and right chiral rotations, namely,
\beq
{\rm SU}(N_f)_L\times {\rm SU}(N_f)_R \times {\rm U}(1)_R,
\label{hesym}
\eeq
where $U(1)_R$ is the nonanomalous  with respect to the non-Abelian gauge bosons $R$ symmetry 
\cite{Sdual} \footnote{The gauge group of our original theory is U$(N)$, thus it includes 
Abelian U(1) gauge fields.  $U(1)_R$ symmetry is anomalous with respect to U(1) gauge fields.
Still we have a freedom to make the U(1) gauge coupling small so the  $U(1)_R$ current is approximately
conserved.}. 
Note, that here we used the fact that $\mu$ is
large and the adjoint matter is decoupled. Say, in the \ntwo limit in which the adjoint matter is 
present the chiral group in (\ref{hesym})  is broken by the Yukawa couplings to the adjoint matter even at small values 
of the quark masses.

The general prescription of the anomaly matching is as follows: the  anomalies of all unbroken global
currents must be the same at all energies well above $m$ (below $m$ chiral symmetries are broken).
In particular, we calculate the anomalies in the ultraviolet (UV) domain in terms of quarks and gauge bosons of 
the original theory and match them with the anomalies calculated in the infrared (IR) domain in terms of the relevant
degrees of freedom of the dual theory. The UV energy should be large enough to ensure
 the original theory to be at weak coupling, $E_{UV}\gg \Lambda_{{\mathcal N}=1}$. Note that 
$\mu$ should be even larger, $\mu\gg E_{UV}$, so the the adjoint matter really decouples and we do have 
chiral symmetry. This explains why we do not check the anomaly matching at intermediate values of $\mu$ (see
Sec. \ref{intermediatemu}).

Under the symmetry (\ref{hesym}) the squark fields transform as 
\cite{Sdual,IS}
\beq
q:\,\left(N_f,\,1,\,\frac{\tN}{N_f}\right),
\qquad
\tilde{q}:\,\left(1,\,\bar{N}_f,\,\frac{\tN}{N_f}\right).
\label{UVq}
\eeq
In particular, the $R$-charges of the squarks under U(1)$_R$ are determined by the number of flavors $N_f$ and 
the rank of the gauge group $N$.
Note, that the fermions (quarks) has $R$-charges $R-1$, where $R$ is the charge of the boson component of a given multiplet,
while gauginos have the  unit $R$ charge.

Quark-like dyons of the $r$-dual theory transform as 
\beq 
D:\,\,\,\left(\bar{N}_f,\, 1,\,\frac{N}{N_f}\right),
\qquad\,\,\,
\tilde{D}:\,\,\, \left(1, \,N_f,\,\frac{N}{N_f}\right),
\label{UVD}
\eeq
where the $R$-charges of dyons are determined by $N_f$ and $\tN$, the rank of the dual gauge group.
Also, in  much the same way as in  \cite{Sdual} we assume that $D$ is in the anti-fundamental representation
of SU$(N_f)_L$. We $\mu$-deform our $r$-dual theory starting from the \ntwo 
limit in which the chiral symmetries are broken. Hence, 
no memory remains as to which of the two SU$(N_f)$ factors in (\ref{hesym}) 
was left-handed or right-handed  in the original quark theory. It is possible that the dyon 
appears  in the fundamental representation  of  SU$(N_f)_R$ at large $\mu$. 
Then the transformations in (\ref{UVD})
ensure (upon redefinition of $D$ and $\tilde{D}$).

The anomaly matching in the IR domain $E_{IR}\gg \sqrt{\xi^{\rm small}}$ closely follows the calculation
in \cite{Sdual}, and we skip it here. The main result is that without the $M$ meson the anomalies do 
not match. Including the $M$ meson we see that it has quantum numbers of $\tilde{q}_Aq^B$ and transforms as \cite{Sdual}
\beq
M:\,\,\,\left(N_f,\,\bar{N}_f,\,\frac{2\tN}{N_f}\right).
\label{UVM}
\eeq
Thus, the anomaly matching requires the presence of the $M$ meson. 

So far we considered the anomaly matching at 
energies $E_{IR}\gg \sqrt{\xi^{\rm small}}$  which ensures that the $M$ meson cannot be heavier than 
$\sqrt{\xi^{\rm small}}$. Below we will show that in fact the upper bound on the the $M$ meson mass is much 
more restrictive.

To this end let us consider energies  $E_{IR}\ll \sqrt{\xi^{\rm small}}$ still well above the scale of the chiral symmetry breaking.
At these energies the unbroken global group is 
\beq
 {\rm SU}(N)\times  {\rm SU}(\tN)\times {\rm U}(1)_V\times {\rm U}(1)_{R'}\,,
\label{IRsym}
\eeq
where first three factors are vector-like symmetries (\ref{c+fd}), while the additional $R$ symmetry
appears in the chiral limit. 

Let us check  that we have an unbroken $R$ symmetry. Consider first dyons of the $r$-dual theory. 
We can combine the $U(1)_R$ transformation with the axial subgroup of the non-Abelian factors in (\ref{hesym})
to make the $R'$ charges of the last $\tN$ dyons  vanish. In this way we arrive at
\beq
R_D'=\frac{N}{N_f} + \left(\frac{\tN}{N_f},...,\frac{\tN}{N_f},-\frac{N}{N_f},...,-\frac{N}{N_f}\right)
=(1,...,1,0,...,0),
\label{Dcomb}
\eeq
where we divide the charges of $N_f$ dyons into $N+\tN$ entries shown in the brackets.
This $U(1)_{R'}$ symmetry is unbroken by the dyon VEVs, see (\ref{Dvev}).

This leads to the following  transformation law of dyons under the unbroken symmetry (\ref{IRsym}):
\beqn
&& 
D^P:\,\,\,\left(\bar{N},\, 1, \,\frac{N_f}{2N},\, 1\right),
\qquad\,\,\,
\tilde{D}_P:\,\,\,\left( N,\, 1, \,-\frac{N_f}{2N},\, 1\right),
\nonumber\\[2mm]
&&
D^K:\,\,\,\left(1,\,\bar{\tN}, \,0,\, 0\right),
\qquad\,\,\,\,\,\,\,\,\,\,\,\,
\tilde{D}_K:\,\,\,\left( 1, \,\tN, \,0,\, 0\right),
\label{IRD}
\eeqn
where $P=1,...,N$ and $K=(N+1),...,N_f$. Here we also combine the vector flavor SU$(N_f)$ transformation
with the $U(1)$ gauge transformation to get vanishing charges  under $U(1)_V$ of the last $\tN$ dyons.

Now let us find the quark $R$ charges. We will see below  that the diagonal entries of
the $N\times N$ upper left  block of the meson 
matrix $M_A^B$ also develops VEVs in the vacuum of the dual theory.
Since the $M$ mesons are defined as quark-antquark pairs of the original theory, this means 
that the $U(1)_{R'}$ symmetry is unbroken if the first $N$ quarks have vanishing $R'$ charges. We define
\beq
R_q'=\frac{\tN}{N_f} + \left(-\frac{\tN}{N_f},...,-\frac{\tN}{N_f},\frac{N}{N_f},...,\frac{N}{N_f}\right)
=(0,...,0,1,...,1).
\label{qcomb}
\eeq 
Thus, the quarks transform under the unbroken symmetry (\ref{IRsym}) as follows:
\beqn
&& 
q^P:\,\,\,\left(N,\, 1, \,0,\, 0\right),
\qquad\,\,\,\,\,\,\,\,\,
\tilde{q}_P:\,\,\,\left( \bar{N},\, 1, \,0,\, 0\right),
\nonumber\\[3mm]
&&
q^K:\,\left(1,\,\tN, \,\frac{N_f}{2\tN},\, 1\right),
\qquad
\tilde{q}_K:\,\left( 1, \,\bar{\tN}, \,-\frac{N_f}{2\tN},\, 1\right),
\label{IRq}
\eeqn
Here we again combine the vector flavor SU$(N_f)$ transformation
with the $U(1)$ gauge transformation to get vanishing charges of the first $N$ quarks under $U(1)_V$.
The transformation properties of the $M$ field ensue from Eq.~(\ref{IRq}),
\beqn
&& 
M^P_{P'}:\,\left(N\bar{N},\, 1, \,0,\, 0\right),
\qquad
M^P_K:\,\left( N,\, \bar{\tN}, \,0,\, 1\right),
\nonumber\\[3mm]
&&
M_P^K:\,\left(\bar{N},\,\tN, \, 0,\, 1\right),
\qquad\,\,
M^K_{K'}:\,\left( 1, \,\tN\bar{\tN}, \,0,\, 2\right),
\label{IRM}
\eeqn
where $P,P'=1,...,N$ and $K,K'=(N+1),...,N_f$.

\vspace{2mm}

The list of anomalies to be checked   is 
\beqn
&&
{\rm U}(1)_{R'}\times  {\rm SU}(N)^2 : \quad-\frac{\delta^{mn}}2\,N|_{UV}= -\frac{\delta^{mn}}2\,N|_{IR},
\rule{0mm}{8mm}\nonumber\\[2mm]
&&
{\rm U}(1)_{R'}\times  {\rm SU}(\tN)^2 :\quad\,\,  0|_{UV}= \frac{\delta^{ps}}2\,(-\tN +\tN)|_{IR},
\nonumber\\[3mm]
&&
{\rm U}(1)_{R'}\times{\rm U}(1)_{V}^2 :\qquad 0|_{UV}= 0|_{IR},
\nonumber\\[4mm]
&&
{\rm U}(1)_{R'}:\qquad -2N^2 +N^2|_{UV} =-N^2=-\tN^2 -N^2 +\tN^2|_{IR},
\nonumber\\[3mm]
&&
{\rm U}(1)_{R'}^3:\qquad -2N^2 +N^2|_{UV} =-N^2=-\tN^2 -N^2 +\tN^2|_{IR},
\nonumber\\[2mm]
\eeqn
where $n,m$ and $p,s$ are the adjoint indices in SU$(N)$ and SU$(\tN)$, respectively.
Here the UV contributions are calculated in terms of the fermion quarks and gauginos, while
the IR contributions come from the fermion components of (screened) dyons and $M$ fields. For example, 
in the second line the
IR anomaly is saturated by $D^K$ and $M^K_{K'}$. In the fourth line the UV contribution comes from 
the quarks $q^P$, $\tilde{q}_P$ and gauginos. The IR contribution comes from the light dyons 
(a half of $D^K$ and $\tilde{D}_K$
states, see Sec.~\ref{UtN}), $M^P_{P'}$ and $M^K_{K'}$, respectively.

Needless to say, all anomalies match. The contribution of the $M$ meson is {\em essential}. Since $E_{IR}$ can lie in the 
window  $m\ll E_{IR}\ll \sqrt{\xi^{\rm small}}$ we find the upper bound for the $M$ meson mass,
\beq
m_{M}\lsim \, m\,.
\label{bound}
\eeq
We see that the $M$ meson is rather light, its mass is determined by the small scale $m$ of the chiral symmetry 
breaking. Thus, the $M$ mesons play a role of $\pi$ mesons in our theory.

\subsection{Interpretation of the Seiberg  \boldmath{$M$} mesons}
\label{interpM}

As was already discussed, the candidates for the Seiberg  $M$ mesons in the $r$-dual theory are stringy mesons
$M_A^B(P,P+1)$ ($P=\tN,...,(N-1)$) from the 
Abelian U(1)$^{(N-\tN)}$ sector. This sector is at  strong coupling at
large $\mu$; therefore,  certain states from this sector can become light. Perturbative states from this sector
(quark-like dyons and Abelian  gauge fields) are singlets with respect to the global group (\ref{IRsym})
and cannot play the $M$ meson role. Note, that stringy mesons $M_A^B(P,P+1)$ (where $P=1,...,(\tN-1)$)
 from the U$(\tN)$ low-energy theory also cannot play the $M$ meson role. First, they are represented
in the U$(\tN)$ low-energy theory by themselves as nonperturbative solitonic states and cannot be added to 
this theory as new ``fundamental'' or ``elementary" fields. Second, they are too heavy, with mass of the order of
$\sqrt{\xi^{\rm small}}$ determined by the tensions of the non-Abelian strings, which can be calculated  
 at weak coupling.

Thus, we propose that the Seiberg $M_A^B$  meson is one  of a multitude of the
monopole-antimonopole stringy mesons
$M_A^B(P,P+1)$ (where $P=\tN,...,(N-1)$) from the Abelian U(1)$^{(N-\tN)}$ sector. At large $\mu$ this
meson should become light, with mass of the order of $m$. It should be
incorporated in the U$(\tN)$ low-energy theory   as a new ``fundamental'' or ``elementary" field. 
Note, that other states from the Abelian sector 
are still heavy and decouple.

\subsection{Effective action}

Since our U$(\tN)$ $r$-dual theory is at weak coupling we can write down its effective action. 
In particular, since this theory is a $\mu$ deformation of a particular \ntwo $r$-dual theory, the  quark-like dyons $D^{lA}$
have  canonically normalized kinetic terms. 
Using
the procedure described in Sec.~\ref{Sconnection} in the opposite direction we ``integrate the $M$-meson
in'' the superpotential (\ref{superpotd}). In this way we arrive at
\beq
{\cal W}= \frac{\kappa^2}{2\mu}\,{\rm Tr}\,(M^2) - \kappa \,m_A\, M_A^A +
\frac{\kappa}{\mu}\,\tilde{D}_{Al}D^{lB}\,M_B^A.
\label{Dsup}
\eeq
We suggest that  (\ref{Dsup}) is a correct continuation of the superpotential (\ref{superpotd})
of the $r$-dual theory to large $\mu$.

Then the effective action of the $r$-dual theory at large $\mu$ takes the form
\beqn
S&=&\int d^4x \left\{\frac1{4\tilde{g}^2}
\left(F^{a}_{\mu\nu}\right)^2 +
\frac1{4\tilde{g}^2_{U(1)}}\left(F_{\mu\nu}\right)^2+
\left|\nabla_{\mu}
D^A\right|^2 + \left|\nabla_{\mu} \bar{\tilde{D}}^A\right|^2
\right.
\nonumber\\[4mm]
&+& \frac2{\gamma}  {\rm Tr}\left|\pt_{\mu} M\right|^2+
\frac{\tilde{g}^2}{2}
\left( 
 \bar{D}_A\,T^a D^A -
\tilde{D}_A T^a\,\bar{\tilde{D}}^A\right)^2
\nonumber\\[3mm]
&+& \frac{\tilde{g}^2_{U(1)}}{8}
\left(\bar{D}_A D^A - \tilde{D}_A \bar{\tilde{D}}^A \right)^2+
 \frac{\kappa^2}{\mu^2}\,{\rm Tr}|DM|^2 +  \frac{\kappa^2}{\mu^2}\,{\rm Tr}|\bar{\tilde{D}}M|^2
\nonumber\\[3mm]
&+& 
\left.
\frac{\gamma}{2}\,\frac{\kappa^2}{\mu^2}\,\left|\tilde{D}_A D^B - \mu m_A \delta_A^B
+\kappa M_A^B\right|^2 
\right\}
\,,
\label{mmodel}
\eeqn
where the
covariant derivative is defined as 
\beq
\nabla_{\mu}= \pt_{\mu} -\frac{i}{2}A_{\mu} - iT^a A_{\mu}^a\,,
\eeq
and we introduced gauge potentials  for   
SU$(\tN)$ and U(1) gauge groups while,  $\tilde{g}$ and $\tilde{g}_{U(1)}$  are associated 
dual gauge couplings. We also introduced the coupling constant $\gamma$ for the $M$ field.

We  assume that $\kappa$ is a function of $\mu$ and $m$ with
the following behavior
\beq
\kappa \sim
\left\{
\begin{array}{l}\rule{0mm}{5mm}
 \mu^{\frac34}\Lambda_{{\mathcal N}=2}^{\frac14} \, , \qquad\;  
 \mu\ll \Lambda_{{\mathcal N}=2} \,,\\[4mm]
 \sqrt{\mu m} \, , \qquad \mu\gg \Lambda_{{\mathcal N}=2} \,  .
 \end{array}
 \right.
\label{kappamu}
\eeq
This dependence ensures that the $M$ meson is heavy, with mass of order of $\sqrt{\xi^{\rm large}}$ 
at intermediate $\mu$, and becomes light, with mass of order of $m$ at large $\mu$.

Minimization of the potential in (\ref{mmodel}) gives VEVs (\ref{Dvev}) for dyons (see also
(\ref{xirN}), (\ref{roots})), while the $M$-field
VEVs    are 
\beq
{\rm diag}\langle M_A^{B}\rangle =
\frac{\mu}{\kappa}\,\left(m_1,...,m_N, 0,...,0\rule{0mm}{4mm}\right).
\label{Mvev}
\eeq
These VEVs ensure chiral symmetry breaking (\ref{IRsym}) in the (almost) equal mass limit.

Now let us briefly discuss mass spectrum of $r$-dual  theory (\ref{mmodel}). 
Much in the same way as at intermediate $\mu$,
the lightest states are $4N\tN$ bifundamental dyons 
with masses of the order of the quark mass differences $(m_A-m_B)$.
A half ($2\tN^2$) of dyons from the singlet and adjoint representations of SU$(\tN)$ have 
 masses of the order of $m$. Moreover, the $M$ mesons are also light,  with masses of the order of $m$.

Other $\tN^2$ dyonic states together with the
gauge bosons of U$(\tN)$ gauge group  are much heavier, with masses
of the order of $\tilde{g}\sqrt{\xi^{\rm small}}$. In addition,  we have stringy monopole-antimonopole
mesons  $M_A^B(P,P+1)$, where  $P=1,...,(\tN-1)$,
with masses   of the  order of $\sqrt{\xi^{\rm small}}$.

However, now at large $\mu$ all these stringy monopole-antimonopole
mesons   can decay into light Seiberg's $M$ mesons.

\section {Vacua with \boldmath{$r<N_f/2$} at small \boldmath{$\mu$}}
\label{zerovacua}
\setcounter{equation}{0}

Now
consider $r$ vacua with $r< N$ in which the first $r$ quarks develop nonvanishing VEVs in the large-mass limit.
In the classically unbroken U$(N-r)$ pure gauge sector the gauge symmetry 
gets broken through the Seiberg--Witten mechanism \cite{SW1}:
first down to U(1)$^{N-r}$ by the condensation of the adjoint fields and then almost completely by the 
condensation of $(N-r-1)$ monopoles. A single
 U(1) gauge factor survives, though,
because the monopoles are charged only with respect to 
the Cartan generators of the SU$(N-r)$ group.

The presence of this unbroken U(1) factor
in all $r<N$ vacua makes them different from the $r=N$ vacuum: in the latter there are
no long-range forces. 

The low-energy theory in the given $r$ vacuum  has the gauge group 
\beq
{\rm U}(r)\times {\rm U}(1)^{N-r}\,,
\label{legaugegroup}
\eeq
if the quark masses are almost equal. Moreover, 
$N_f$  quarks are charged
under the U$(r)$ factor, while  $(N-r-1)$ monopoles are charged under the U(1) factors.
If $0<r<(N-1)$ then the $r$-vacua are hybrid vacua in which both, quarks and monopoles, are condensed. Note that the
quarks and monopoles are charged with respect to orthogonal subgroups of U$(N)$
and therefore are mutually local (i.e. can be described by a local Lagrangian). 
The low-energy theory
is infrared-free and it is at  weak coupling as long as VEVs of quarks and monopoles are small. 
The quark VEVs are given by
\beqn
\langle q^{kA}\rangle &=& \langle\bar{\tilde{q}}^{kA}\rangle=\frac1{\sqrt{2}}\,
\left(
\begin{array}{cccccc}
\sqrt{\xi_1} & \ldots & 0 & 0 & \ldots & 0\\
\ldots & \ldots & \ldots  & \ldots & \ldots & \ldots\\
0 & \ldots & \sqrt{\xi_r} & 0 & \ldots & 0\\
\end{array}
\right),
\nonumber\\[4mm]
k&=&1,..., r\,,\qquad A=1,...,N_f\, ,
\label{qvevr}
\eeqn
where in the  quasiclassical domain of large quark masses the $r$ parameters $\xi_{1,..., r}$ are
\beq
\xi_P \approx 2\;\mu m_P,
\qquad P=1,..., r\,.
\label{xiclass}
\eeq
These parameters can be made small in the  limit of  large $m_A$ if $\mu$ is sufficiently small.

In quantum theory  the $\xi_P$ parameters   are determined by the roots of the Seiberg-Witten curve
(\ref{curve}), see  \cite{SYrvacua,SYhybrid}.
The Seiberg-Witten curve in the $r<N$ vacuum has $N-1$ double roots which are 
 associated with $r$ condensed quarks and $(N-r-1)$ condensed monopoles.

Namely, the  Seiberg--Witten curve factorizes \cite{CaInVa},
\beq
y^2
=\prod_{P=1}^{r} (x-e_P)^2\,\prod_{K=r+1}^{N-1} (x-e_K)^2\,(x-e_N^{+})(x-e_N^{-})\,.
\label{rcurve}
\eeq
The first $r$ quark double roots are associated with  the
mass parameters in the large mass limit, $\sqrt{2}e_P\approx  -m_P$, where $P=1, ... , r$. 
The other $(N-r-1)$ double roots  associated with the light monopoles  are much smaller, 
and are determined by $\Lambda_{{\mathcal N}=2}$.
The last two  unpaired roots  are also much smaller. 
 For the single-trace deformation superpotential (\ref{msuperpotbr}) 
 their sum vanishes \cite{CaInVa},
\beq
e_N^{+} + e_N^{-}=0\,.
\label{DijVafa}
\eeq
The root $e_N^{+}$ determines the value of the gaugino condensate \cite{Cachazo2},
\beq
e_N^2=\frac{2S}{\mu}, \qquad S=\frac1{32\pi^2}\langle {\rm Tr}\,W_{\alpha}W^{\alpha} \rangle\,.
\label{eN}
\eeq
The superfield $W_{\alpha}$ includes the gauge field strength tensor.

In terms of the roots of the Seiberg-Witten curve the quark VEVs  are given by 
the formula \cite{SYrvacua,SYhybrid}
\beq
\xi_P=-2\sqrt{2}\,\mu\,\sqrt{(e_P-e_N^{+})(e_P-e_N^{-})}, \qquad P=1,...,(N-1).
\label{xi}
\eeq
In fact, this formula is universal: it determines both, the VEVs of $r$ quarks and $(N-r-1)$ monopoles
\cite{SYhybrid}. Namely, the index $P$ runs over $P=1,...,(N-1)$ in (\ref{xi}) with quark and monopole VEVs 
given by (\ref{qvevr}) and 
\beq
  \langle m_{P(P+1)}\rangle = \langle\bar{\tilde{m}}_{P(P+1)}\rangle = \sqrt{\frac{\xi_P}{2}},
\qquad  P=(r+1),..., (N-1) ,
 \label{mvev}
 \eeq
respectively.
 Here $m_{PP'}$ denotes the  monopole with the charge given by  
the root $\alpha_{PP'}=w_P-w_{P'}$ of the  SU$(N)$ algebra with the weights $w_P$ ($P<P'$).

Condensation of $r$ quarks leads to formation of non-Abelian magnetic strings that confine
monopoles from the SU($r$) sector (strings are non-Abelian in the (almost) equal quark mass limit). 
Tensions of the magnetic strings are determined by (\ref{ten}) with $P=1,...,r$. 
In a similar way condensation of $(N-r-1)$ monopoles leads to the formation of the Abelian electric strings
which confine quarks from U(1)$^{N-r}$. Their tensions are also given by Eq.~(\ref{ten}) 
with $P=(r+1),...,(N-1)$, for more details on confinement of monopoles and quarks in the hybrid vacua see 
\cite{SYhybrid}.

Now let us consider the limit of small quark masses. As was already mentioned, in the $r$ vacua with $r>N_f/2$
there is a crossover to   the  $r$-dual theory with the dual gauge group U$(N_f-r)\times$U(1)$^{N-N_f+r}$
\cite{SYrvacua}. The $r=N$ vacuum considered in the previous sections provides us with the simplest example of 
this behavior. 

Now let us focus on $r$-vacua with smaller $r$. If  $r< N_f/2 $ 
 the low-energy theory essentially remains the same as at
 large $m_A$, namely,
infrared-free U$(r)\times$U(1)$^{N-r}$ gauge theory with $N_f$ flavors of light states charged under 
non-Abelian gauge factor and 
 $(N-r-1)$ singlet monopoles charged under U(1)$^{N-r}$ \cite{MY2,SYzerovac}.
 Although the color charges of light 
non-Abelian
states are identical to those of quarks\,\footnote{ As we reduce  $m$  the
quarks pick up root-like color-magnetic charges, in addition to their weight-like color-electric charges
 due to monodromies, see \cite{MY2}.} they are not quarks. In much the same way as in the $r=N$ vacuum
we call these states quark-like dyons $D^{lA}$, $l=1,...,r$, $\, A=1,...,N_f$. We will see in 
Sec.~\ref{largemu0} that they have chiral $R$-charges different from those of quarks.\footnote{In \cite{SYzerovac}
the chiral limit was not considered. It was concluded that these states are identical to quarks. Here we correct
this interpretation.} At large masses these dyons are heavy monopole-antimonopole stringy  states
 while below crossover, at small masses, 
they become light fundamental (or elementary) states of the U$(r)\times$U(1)$^{N-r}$ gauge theory.

The quark-like dyons from the U$(r)$ sector and the monopoles from the 
orthogonal U(1)$^{N-r}$ sector  develop VEVs 
determined by Eq. (\ref{xi}). In particular, dyons develop VEVs
\beqn
\langle D^{lA}\rangle &=& \langle\bar{\tilde{D}}^{lA}\rangle=\frac1{\sqrt{2}}\,
\left(
\begin{array}{cccccc}
\sqrt{\xi_1} & \ldots & 0 & 0 & \ldots & 0\\
\ldots & \ldots & \ldots  & \ldots & \ldots & \ldots\\
0 & \ldots & \sqrt{\xi_r} & 0 & \ldots & 0\\
\end{array}
\right),
\nonumber\\[4mm]
l&=&1,..., r\,,\qquad A=1,...,N_f\, .
\label{Dvevr}
\eeqn
The theory   is at weak coupling provided the $\xi_P$  parameters   are small.

What happens to quarks of the original theory? In much the same way 
as in the $r=N$ vacuum the screened $q^{kA}$ quarks (with $k=1,...,r$)
of the U$(r)$ gauge sector  decay into monopole-antimonopole pairs and evolve into stringy
mesons shown in Fig.~\ref{figmeson}. These quarks are in the instead-of-confinement phase.

We would like to stress however, that there is a peculiar distinction of this picture with the one 
in the $r=N$ vacuum. In the limit of small and almost equal masses the dyon condensation breaks
the global SU$(N_f)$ group down to 
\beq
 {\rm SU}(r)_{C+F}\times  {\rm SU}(N_f-r)\times {\rm U}(1)_V\,.
\label{c+f0vac}
\eeq
In particular, color-flavor locking takes place in the SU$(r)$ factor. In contrast to the case of 
the $r=N$ vacuum
both dyons and monopole-antimonopole stringy mesons, which originate from screened quarks of the large-$m$
theory are in the same representations of this group. Namely, they form singlet and adjoint representaions
of SU$(r)_{C+F}$ as well as bifundamental representations,
\beq
(1,\, 1), \quad (r^2-1,\, 1),
\quad (\bar{r},\, N_f-r), \quad
(r,\, \bar{N}_f-\bar{r})\,,
\label{repres0vac}
\eeq
where we    mark representations with respect to two 
non-Abelian factors in (\ref{c+f0vac}). The U(1)$_R$ symmetry which distinguishes screened dyons and 
monopole-antimonopole  mesons (former screened quarks) is broken. Therefore 
monopole-antimonopole stringy mesons are unstable and  decay into  dyons, which are lighter.

There are also other quarks $q^{kA}$ charged with respect to the Abelian U(1)$^{N-r}$ gauge group
with $l=(r+1),..., N$ in the original theory. These are still confined by Abelian strings 
formed as a result of  the monopole condensation in the small-$m$ limit.

\section {Zero vacua}
\label{0vacuamu}
\setcounter{equation}{0}

In this section we consider  zero vacua at intermediate and large  $\mu$  \cite{SYzerovac}.
 These vacua  form a subset of $r$ vacua with small $r$, $r<\tN$.

\subsection{Intermediate \boldmath{$\mu$}}

In the small mass limit $r$  double roots of the Seiberg-Witten curve associated with light dyons
are still determined by quark masses
\beq
\sqrt{2}e_P = -m_{P},\qquad P=1,...,r\,.
\label{rroots}
\eeq
The above expression is valid in $r$ vacua with $r<N_f/2$. Other roots are much larger, of the order of $\Lambda_{{\mathcal N}=2}$. 
However, in contrast to the $r=N$ vacuum (see Sec.~\ref{UtN}) this 
 does not allow us to increase $\mu$ keeping the U$(r)$ theory at weak coupling. The point is that
dyons' VEVs which are supposed to be small to ensure weak coupling (in the IR free theory) are not determined 
entirely by $e_P$ in the $r<N$ vacua. They are given by parameters $\xi_P$  that depend also on the
gaugino condensate which determines the values of the unpaired roots in (\ref{xi}). In the majority of the $r$ vacua 
the gaugino condensate is of the order of $S\sim \mu\Lambda^2_{{\mathcal N}=2}$. We refer to these vacua 
as the $\Lambda$ vacua.
In the $\Lambda$ vacua all parameters $\xi$ are of the order of $\xi\sim \mu\Lambda_{{\mathcal N}=2}$, and we cannot
increase $\mu$ without destroying the weak coupling condition \cite{SYzerovac}.

However, there are two exceptions. One is the $r=N$ vacuum in which the gaugino condensate vanishes, and 
 $\tN$ parameters $\xi$ are determined  by the quark  masses, see (\ref{xirN}) and (\ref{roots}) \cite{SYN1dual}.
We considered this vacuum in the previous sections.
Another exception is the subset of the $r<\tN$ vacua, which we call the zero vacua \cite{SYzerovac}. In the zero vacua the
gaugino condensate is extremely small \cite{GivKut,SYzerovac},
\beq
S\approx  \mu\, 
\frac{m^{\frac{N_f-2r}{\tN -r}}}{\Lambda_{{\mathcal N}=2}^{\frac{N-\tN}{\tN-r}}}
\;e^{\frac{2\pi k}{\tN-r}\,i}\, \ll \mu m^2, 
\qquad 
 k=1,... , (\tN-r)\,,
\label{0vacgaugino}
\eeq
in the limit of small equal quark masses. This behavior can be obtained from the
exact Cachazo-Seiberg-Witten solution for the chiral ring of the theory \cite{Cachazo2}, see also \cite{SYzerovac}.

Thus in the zero vacua we can neglect contributions of the unpaired roots  as compared to the quark masses in 
(\ref{xi}). It turns out that $\xi$'s are given by \cite{SYzerovac}
\beq
\xi_P \approx -2\mu \,\left(m_1,...,m_r,\,0,...,0,\,\Lambda_{{\mathcal N}=2},...,
\Lambda_{{\mathcal N}=2}e^{\frac{2\pi i}{N-\tN}(N-\tN-1)}\right),
\label{xi0vac}
\eeq
where $(\tN-r)$ entries are of the order of $\sqrt{S/\mu}$ and taken to be zero in the quasiclassical 
approximation,
while the last entries are large. They determine VEVs of $(N-\tN)$ monopoles.

 Now we can  increase $\mu$ to intermediate values
\beq
|\mu| \gg |m_A |, \qquad A=1, ... , N_f\, , \qquad \mu\ll \Lambda_{{\mathcal N}=2} .
\label{muintermed0}
\eeq
The monopole U(1)$^{(\tN-r)}$ sector associated with almost vanishing entries in (\ref{xi0vac}) enters the strong
regime. It is shown in \cite{SYzerovac} that it goes through a crossover at $\mu\sim e_N\sim \sqrt{S/\mu}$, 
and 
the domain of intermediate $\mu$ can be described in terms of  the weak coupling
 $\mu$ dual theory with the gauge group 
\beq
{\rm U}(\tN)\times {\rm U}(1)^{N-\tN}\,,
\label{mudualgaugegroup}
\eeq
$N_f$ flavors of quark-like dyons  charged with respect to the U$(\tN)$ gauge group and $(N-\tN)$ singlet 
monopoles 
charged with respect to  the U(1)$^{N-\tN}$  Abelian sector. The restoration of the U$(\tN)$ gauge group occurs 
 because
$(\tN-r)$ Coulomb branch parameters $\phi_P$ of the Seiberg-Witten curve  almost vanish, being determined 
by the small value of the gaugino condensate \cite{SYzerovac}.

Qualitatively the enhancement of the U$(r)$ gauge group to U$(\tN)$ can be understood as follows. As we reduce $m$, the expectation values  of monopoles in the U$(\tN-r)$ sector tend to zero, see (\ref{xi0vac}). Confinement of quarks in this sector 
becomes weaker and eventually disappears. However, confined quark-antiquark pairs cannot just move apart
because they have ``wrong'' chiral charges, see the next subsection. They decay into a pair of quark-like dyons
\beq
q+\tilde{q} \to \bar{D} + \bar{\tilde{D}} + \lambda +\lambda
\label{decay}
\eeq 
via emission of two gauginos.

These dyons and gauge fields of the U$(\tN-r)$ sector become unconfined and enter the non-Abelian Coulomb phase.
Moreover, dyons  of the U$(\tN-r)$ sector combine with dyons of the U$(r)$ sector to form light non-Abelian matter 
of the enhanced U$(\tN)$ gauge group.

Note also that VEVs of $r$ dyons  are given by $\xi^{\rm small}$ while VEVs of $(N-\tN)$ monopoles
 are much larger and given by
$\xi^{\rm large}$, see (\ref{xi0vac}). Therefore, the monopole sector is heavy and can be integrated out
together with the adjoint matter. In much the same way as in the $r=N$ vacuum this leaves us with the 
low-energy theory with Seiberg's dual gauge group 
\beq
{\rm U}(\tN)\,
\label{gg0vac}
\eeq
  and $N_f$ flavors of dyons  with the superpotential  \cite{SYzerovac}
\beq
{\mathcal W}_{\rm zero\; vac} = -\frac1{2\mu}\,
(\tilde{D}_A D^B)(\tilde{D}_B D^A)  
+m_A\,(\tilde{D}_A D^A)\,.
\label{sup0vacintmu}
\eeq
This is the same superpotential as in the $r=N$ vacuum, see (\ref{superpotd}).

Note, that the dyons in this setup have  $\tN$ colors, however, only $r$ of them condense, $r<\tN$. 
Thus our low-energy  infrared free U$(\tN)$
theory is in the mixed Coulomb-Higgs phase with regards to dyons.  In particular, the U$(\tN-r)$ subgroup of 
U$(\tN)$
remains unbroken, and $(\tN-r)$ massless gauge bosons are present. The gauge bosons of the U$(r)$ subgroup
and their dyon \none superpartners have  masses of the order of $\tilde{g}\sqrt{\xi^{\rm small}}$. Other
dyons charged with respect to  U$(\tN)$ have masses of the order of $m$.

\vspace{2mm}

Quarks of the original theory charged with respect to U(1)$^{N-\tN}$ are confined by
electric strings formed due to the condensation of monopoles in the heavy U(1)$^{N-\tN}$  Abelian sector. 
In much the similar way as in the $r=N$ vacuum these stringy mesons are the candidates for Seiberg's
$M$ mesons. At intermediate values of $\mu$ the U(1)$^{N-\tN}$  Abelian sector is at weak coupling, and  these 
mesons are heavy, with masses of the order of $\sqrt{\xi^{\rm large}}\sim \sqrt{\mu \Lambda_{{\mathcal N}=2}}$.

We can compare our low-energy  U$(\tN)$ $\mu$-dual theory to Seiberg's dual. In much the same way as in the
$r=N$ vacuum we find a perfect match \cite{SYzerovac}. Namely, if we integrate out $M$ fields
in Seiberg's dual superpotential (\ref{Ssup}) (they are heavy at intermediate values of $\mu$) and make 
identification  (\ref{change})
similar to that in the $r=N$ vacuum  we arrive at the superpotential which coincides (up to a sign) with our
superpotential (\ref{sup0vacintmu}).

The identification (\ref{change}) reveals the physical nature of the Seiberg ``dual quarks.''
In much the same way as in the $r=N$ vacuum they are not monopoles. Instead, they are quark-like dyons 
which have color charges identical to those of quarks but different global charges. Condensation of $r$
dyons leads to confinement of monopoles and the ``instead-of-confinement" phase for quarks in the U$(r)$ sector.

\subsection{Large \boldmath{$\mu$}}
\label{largemu0}

Now we assume that $\mu$ is large while $\sqrt{\xi^{\rm small}}$ is small enough to ensure the weak coupling
regime in the low-energy  U$(\tN)$ $\mu$-dual theory, see (\ref{mularge}). By the same token as in 
the $r=N$ vacuum we can use the anomaly matching to show that Seiberg's $M$ mesons should become light 
at large $\mu$. 

If the IR energy scale is large, $E_{IR} \gg\sqrt{\xi^{\rm small}}$, the global group
is given by (\ref{hesym}) and in this case the anomaly matching was carried out in \cite{Sdual}.
Namely, the transformation properties of quarks of the original theory and $M$ mesons are given by
Eqs. (\ref{UVq}) and (\ref{UVM}). Let us consider the dyon charges. The $R$ charge is determined by the
anomaly cancellation requirement with respect to  non-Abelian  gauge bosons \cite{Sdual}. It
is determined by the number of flavors and the rank of the gauge group. Say, for quarks of the original
 theory it is  given by $\tN/N_f$, see (\ref{UVq}). The rank of the gauge group in the $\mu$-dual theory 
is different, however. It   equals  $\tN$. Thus, the $R$ charges of the $D^{lA}$ dyons  are given by
\beq
R_{D}=\frac{N}{N_f}\,.
\label{RchargeD}
\eeq

This tells us that the quarks and dyons are in fact different states, as was mentioned above. We arrive at our
$\mu$-dual theory starting from the \ntwo limit by virtue of the $\mu$ deformation. Moving along this way 
we break the U(1)$_R$ symmetry. Thus, we were unable to observe the above distinction. The dyons appeared just as  quarks with a
truncated number of colors. Now, studying the chiral limit, we see that
in fact they are different states. 

As was already explained, the  weakly confined quark-antiquark pairs decay
into unconfined dyon pairs via a wall-crossing-like process 
\beq
q+\tilde{q} \to \bar{D} + \bar{\tilde{D}} + \lambda +\lambda,
\label{decay1}
\eeq 
upon increasing $\mu$, see (\ref{decay}).
It is easy to see that this decay respects the $R$-charge conservation, where we use the fact that the gaugino $R$ charge  
 is unity. Equation~(\ref{decay1}) shows that the dyon transformation laws  are 
\beq 
D:\,\,\, \left(\bar{N}_f,\, 1,\,\frac{N}{N_f}\right),
\qquad
\tilde{D}:\,\,\,\left(1, \,N_f,\,\frac{N}{N_f}\right).
\label{UVD0vac}
\eeq
In particular, the $D^{lA}$ dyon  transforms in the $\bar{N}_f$ representation of the SU$(N_f)_L$ rather 
than\,\footnote{This important  circumstance was  noted 
by Chernyak \cite{Chernyak:2013tsa}.} in the representation $N_f$.

We see that the dyon transformation properties   are the same in both, the 
zero and $r=N$ vacua (see (\ref{UVD})), and coincide with those for the Seiberg dual quarks \cite{Sdual}.
Thus, the anomaly matching at the IR energy scale $$E_{IR} \gg\sqrt{\xi^{\rm small}}$$ follows the calculation
presented in  \cite{Sdual}.
 The concluding result is: the   light neutral $M_A^B$ field  is needed to match the  anomalies.

If  $E_{IR} \ll\sqrt{\xi^{\rm small}}$, the unbroken global group is 
\beq
 {\rm SU}(r)\times  {\rm SU}(N_f-r)\times {\rm U}(1)_V\,.
\label{IRsym0vac}
\eeq
In particular, it is easy to see that the  chiral  U(1)$_R$ symmetry is broken in the zero vacua
in contradistinction with the  $r=N$ case. In fact, we cannot arrange combinations similar to that in (\ref{Dcomb}) and (\ref{qcomb})
to ensure that the $R'$ charges of $r$ components of quarks and $(N_f-r)$ components of $M$ mesons (which develop VEVs) vanish.
The required axial rotation from the non-Abelian subgroups in (\ref{hesym}) does not respect the 
Yukawa interaction $(\tilde{D}_A D^B)M_B^A$. Therefore, we cannot match anomalies at energies below
$\sqrt{\xi^{\rm small}}$.

Thus, in the zero vacua the anomaly matching gives a less restrictive upper bound on the $M$-meson mass   
 as compared to the
$r=N$ vacuum, namely $m_M\lsim \sqrt{\mu m}$. Still we can obtain a more restrictive estimate 
for the $M$-meson mass using the Goldstone theorem. The number of broken generators in the breaking
of (\ref{hesym}) down to (\ref{IRsym0vac}) is
\beq
r^2 +(N_f-r)^2 + 4r(N_f-r).
\label{brokgen}
\eeq
While $r^2$ and $4r(N_f-r)$ broken generators can be accounted for by light dyons in the $r\bar{r}$ and 
bifundamental representations, respectively, the extra $(N_f-r)^2$ light states are missing. 
These can be accounted for by the light $M$ meson. As a result, we conclude that $M$-meson mass should be
lighter, namely
\beq
m_M\sim  m\,,
\label{bound0vac}
\eeq
as is the case in the $r=N$ vacuum.

The physical interpretation of the Seiberg's $M$ mesons in the zero vacua is as follows. As was already mentioned, 
the candidates for the $M$ mesons can be found among mesonic states from the heavy Abelian U(1)$^{N-\tN}$ sector --
quark-antiquark pairs connected by confining strings. The majority of these mesons are similar to 
those shown in Fig.~\ref{figmeson} in which the monopoles should be replaced by quarks. However, a peculiar feature
of all $r<N$ vacua is that there are only $(N-1)$ strings, one of strings is missing. Therefore,
some of these mesons are formed by quarks and antiquarks connected by only one string, while the other one
is missing,
see \cite{SYrvacua,SYhybrid} for more details.

Now, similarly to the situation in
the $r=N$ vacuum, we suggest that one of these quark-antiquark stringy meson become light at large
$\mu$ when the U(1)$^{N-\tN}$ sector enters the strong coupling regime. This $M$ meson should be integrated in 
the U$(\tN)$ $\mu$-dual theory as a ``fundamental'' (elementary) field. Other fields of the Abelian U(1)$^{N-\tN}$ sector
are heavy and can be integrated out. The superpotential and action of the low-energy U$(\tN)$ $\mu$-dual theory
are given in Eqs.  (\ref{Dsup}) and (\ref{mmodel}).

\section {Summary and Conclusions}
\label{conclusion}
\setcounter{equation}{0}

To summarize, at large $\mu$ and small $\xi^{\rm small}$  $\mu$-deformed SQCD in the $r=N$ vacuum 
 is described by the 
weakly coupled infrared-free  $r$-dual  U$(\tN)$ theory (\ref{mmodel}) with   
$N_f$ light quark-like dyon flavors. Condensation of the light dyons
$D^{lA}$ in this theory triggers formation of the non-Abelian strings and confinement of monopoles.
The quarks and gauge bosons
 of the original \none SQCD are in the ``instead-of-confinement'' phase: they evolve  into
 monopole-antimonopole stringy mesons shown in Fig.~\ref{figmeson}. There is also  Seiberg's 
neutral meson $M$ field which is monopole-antimonopole stringy meson from heavy Abelian sector. It becomes
anomalously light and plays the role of a ``pion'' at large $\mu$.

In the zero $r$-vacua we  have the weak coupling description in terms of the
infrared-free  $\mu$-dual  U$(\tN)$ theory
(\ref{mmodel}) with $N_f$ flavors of quark-like dyons. Only $r$ dyons condense ($r<\tN$) 
leading to  confinement 
of monopoles in the U$(r)$ sector. The U$(\tN-r)$ sector is in the non-Abelian Coulomb phase for dyons. 
 The Seiberg's $M$-meson is a quark-antiquark stringy state which comes from
the heavy Abelian sector. It becomes light at large $\mu$.

\section*{Acknowledgments}
This work  is supported in part by DOE grant DE-FG02-94ER40823. 
The work of A.Y. was  supported 
by  FTPI, University of Minnesota, 
by RFBR Grant No. 13-02-00042a 
and by Russian State Grant for 
Scientific Schools RSGSS-657512010.2.

\end{document}